\documentclass[doublespacing]{bmcart}

\usepackage[margin=1cm]{caption}
\usepackage{endfloat}
\usepackage{amsfonts}
\usepackage{graphicx}

\usepackage{graphicx}

\usepackage{subcaption}

\usepackage{amssymb}
\usepackage{listings}

\usepackage{amsthm}
\usepackage{latexsym}
\usepackage{xspace}
\usepackage{amsmath}
\usepackage{epsfig}
\usepackage{mathrsfs}
\usepackage{url}

\newtheorem{ex}{Example}

\newcommand{\bone}{\boldsymbol{1}}

\newcommand{\bx}{\boldsymbol{x}}

\newcommand{\bm}{\boldsymbol{m}}
\newcommand{\by}{\boldsymbol{y}}

\newcommand{\bg}{\boldsymbol{g}}

\newcommand{\be}{\boldsymbol{e}}
\newcommand{\bzero}{\boldsymbol{0}}

\usepackage[utf8]{inputenc} 

\def\includegraphics{}

\startlocaldefs
\endlocaldefs

\begin{document}

\begin{frontmatter}

\begin{fmbox}
\dochead{Research}


\title{Training population selection for  (breeding value) prediction}


\author[
   addressref={aff1},                   
   corref={aff1},                       
   email={da346@cornell.edu}   
]{\inits{DA}\fnm{Deniz} \snm{Akdemir}}


\address[id=aff1]{
  \orgname{Department of Plant Breeding \& Genetics, Cornell University}, 
  \city{Ithaca, NY},                              
  \cny{USA}                                    
}


\end{fmbox}


\begin{abstractbox}

\begin{abstract} 
Training population selection for genomic selection has captured a great deal of interest in animal and plant breeding. In this article, we derive a computationally efficient statistic to measure the reliability of estimates of genetic breeding values for a fixed set of genotypes based on a given training set of genotypes and phenotypes. We adopt a genetic algorithm scheme to find a training set of certain size from a larger set of candidate genotypes that optimizes this reliability measure. Our results show that, compared to a random sample of the same size, phenotyping individuals selected by our method results in models with better accuracies. We implement the proposed training selection methodology on four data sets, namely, the arabidopsis, wheat, rice and the maize data sets. Our results indicate that dynamic model building process which uses genotypes of the individuals in the test sample into account while selecting the training individuals improves the performance of GS models.
\end{abstract}




\end{abstractbox}
%

\end{frontmatter}




\section*{Introduction}

Breeding through genomic selection (GS) in animal or plant breeding is based on estimates of genetic breeding values (GEBVs). Prediction of the GEBVs usually involves implementing a whole-genome regression model where the known phenotypes are regressed on the markers. In GS, first a set of genotypes to be phenotyped (a training population) are identified and phenotyped. Once the phenotypes are measures for the training set of individuals, a regression model is trained to predict GEBVs for individuals which were not phenotyped. Finally, these GEBVs are used for evaluation of individuals. Since phenotyping is a time consuming and costly process selecting a good training population is essential for the success of GS.

In this article we concentrate on the first step of GS, i.e., the selection of training population, to address the accuracy of the GS models. We imagine a scenario in which we are given two sets of individuals and their markers. The first set includes the candidate individuals from which a training set is to be selected for phenotyping to predict the GEBVs of the individuals in the second test set. It will be shown that a model building process which uses genotypes of the individuals in the test sample into account while selecting the training individuals improves the performance of prediction models.

Various regression models have been successfully used for predicting the breeding values in plants and animals. In both simulation studies and in empirical studies of dairy cattle, mice and in bi-parental populations of maize, barley and \emph{Arabidopsis} marker based GEBVs have been quite accurate. However, it has also been shown that as the training and testing population diverge the accuracies of the GEBVs decrease. As the breeding populations tend to change over time, the result is that the accuracies of the GEBVs obtained from the training population decrease over time. Similarly, in the existence of strong population structure, the GEBVs obtained by using sub-populations are usually not accurate for individuals in other sub-populations.  

In breeding, the problem of training population selection has captured some attention. For example, the relaibility measure of VanRaden (\cite{vanraden2008}) is expressed as \begin{equation}\label{VanRaden}K_{21}(K_{11}+\delta I)^{-1}K'_{21}\end{equation} where $K_{21}$ is the matrix  of genomic relationships between the individuals in the test set to each of the individuals in the training set and $K_{11}$ measures the genomic relationships in the training set and finally the parameter $\delta$ is related to the heritability ($h$) of the trait by $\delta=(1-h^2)/h^2.$ This reliability measure is related to Henderson's prediction error variance (PEV) (\cite{henderson1975best}) and the more recent coefficient of determination (CD) of Laloe (\cite{laloe1996considerations}) which were both utilized in (\cite{rincent2012maximizing}) for the training population selection problem.

The optimization of the reliability measure in \ref{VanRaden} and the related PEV and CD require expensive evaluations (inversion of large matrices) many times therefore they are not computationally feasible for large applications. In the next sections, we derive a computationally efficient approximation to the PEV and use this measure for the training population selection. Another novelty in our method compared to the optimization schemes recommended in (\cite{rincent2012maximizing}) is that in our case we calculate the prediction error variance for the individuals in the test set instead of evaluating it within the candidate set, i.e., we use domain information about the test data while building the estimation model by selecting in the individuals to the training set such that they minimize the PEV in the test set. The methods developed here can be used for dynamic the model building, in other words, different test sets will amount to different individuals be selected from the candidate set and hence different estimation models.

\section*{Methods}
Traditionally, the breeder is interested in the total additive genetic effects as opposed to the total genetic value. Therefore, a linear model is assumed between the markers and the phenotypes. This is expressed as writing \begin{equation}y=\beta_0+\bm'\beta+e\end{equation} where $y$ stands for the  phenotype, $\beta_0$ is the mean parameter, $\bm$ is the $m-$vector of marker values, $\beta$ is the $m-$vector of marker effects and $e,$ the difference between the observed and the fitted linear relationship, has a normal distribution with zero mean and variance $\sigma_e^2.$  

In order to estimate the parameters of this model, we will acquire $n_{Train}$ individuals from a larger candidate population. The model the will be used to estimate a fixed set of $n_{Test}$ individuals. 

Let M bet the matrix of markers partitioned as \begin{equation}M=\left[
\begin{array}{c}
M_{Candidate}\\ \hline
M_{Test}
\end{array}
\right]\end{equation}
where $M_{Candidate}$ is the $n\times m$ matrix of markers for the individuals in the candidate set and $M_{Test}$ is the matrix of markers for the individuals in the test set. We would like to identify $n_{Train}$ training set individuals from the candidate set (and therefore a matrix $M_{Train}$) for which the average prediction variance for the individuals in the test set needs to be minimized. Given we have determined $M_{Train}$ and observed their phenotypes $\by_{Train},$ we can write \begin{equation}\by_{Train}=(\bone, M_{Train})(\beta_0, \beta')'+\be.\end{equation} Under the assumptions of this model the uniformly minimum variance estimators for the phenotypes in the test data is expressed as \begin{equation}\widehat{\by}_{Test}=(\bone, M_{Test})((\bone,M_{Train})'(\bone,M_{Train}))^{-}(\bone,M_{Train})'\by_{Train}\end{equation} where the $^{-}$ denotes the pseudo inverse of a matrix. Ignoring the constant term, $\sigma_e^2,$ the covariance matrix (Prediction Error Variance (PEV)) for $\widehat{\by}_{Test}$ is \begin{equation} PEV(M_{Test})=(\bone, M_{Test})((\bone,M_{Train})'(\bone,M_{Train}))^{-}(\bone,M_{Test})'.\end{equation}

With the emergence of modern genotyping technologies the number of markers can vastly exceed the number of individuals. To overcome the problems emerging in these large $m$ with small $n$ regressions, estimation procedures performing variable selection, shrinkage of estimates, or a combination of both are commonly used while estimating the effects of markers. These methods trade the decreasing variance to increasing bias due to shrinkage of individual marker effects to obtain a better overall prediction performance. Since the variance of these selection-shrinkage methods will be smaller than the least squares estimators, the $PEV(M_{Test})$ is an upper bound on the covariance matrix of the PEV of these models.  To see this consider the PEV from the ridge regression: \begin{equation}\label{PEVRIDGE}PEV^{Ridge}(M_{Test})=(\bone, M_{Test})((\bone,M_{Train})'(\bone,M_{Train})+\lambda I)^{-1}(\bone,M_{Test})'\end{equation}. Clearly, $PEV^{Ridge}(M_{Test})\leq PEV(M_{Test})$ for any $\lambda\geq 0.$

We would like to obtain minimum variance for our predictions in the test data set. Therefore, we recommend minimizing \begin{equation}tr(PEV(M_{Test}))\end{equation} with respect to $M_{Train}$ when selecting individuals to the training set. 

The training data evaluation criterion $PEV$ is related to the integrated average prediction variance (IV), where \begin{equation} IV=\frac{1}{A}\int_\chi\bx'(X'_{Train}X_{Train})^{-1}\bx d\bx\end{equation}
where $A$ is the volume of the space of interest $\chi.$ See Box and Draper (\cite{box1959basis}) for a detailed discussion of this criterion. A design that minimizes IV is referred to as IV-optimal.

However, since we are dealing with a large number of markers and any optimization scheme would involve numerous evaluation of this objective function the formula for the $PEV(M_{Test})$ is not practically applicable. A more suitable numerically efficient approximation to $PEV(M_{Test})$ can be obtained by using the first few principal components (PCs) of the markers matrix $M$ instead of $M$ in the training population selection stage. Let  $P$ be the matrix of first $k\leq min(m, n)$ PCs partitioned as \begin{equation}P=\left[
\begin{array}{c}
P_{Candidate}\\ \hline
P_{Test}
\end{array}
\right]\end{equation}
where $P_{Candidate}$ is the matrix of PCs for the individuals in the candidate set and $P_{Test}$ is the matrix of PC's for the individuals in the test set. Now, $PEV^{Ridge}(M_{Test})$ can be approximated by \begin{equation}\label{PEVRIDGEAPPROX}PEV(M_{Test})\approx (\bone, P_{Test})((\bone,P_{Train})'(\bone,P_{Train})+\lambda I)^{-1}(\bone,P_{Test})'.\end{equation}

Finally, we would like to note that the $PEV(M_{Test})$ is related to the reliability measure in (\ref{VanRaden}).  To see this, write \begin{equation}  (M'_{Train}M_{Train}+\lambda I)^{-1}=\frac{1}{\lambda}(I-M'_{Train}(M_{Train}M'_{Train}+\lambda I)^{-1}M_{Train}.\end{equation} 

Lettting $\delta=m\lambda,$ $K_{21}= M_{Test}M'_{Train}/m,$ $K_{11}=$ $ M_{Train}M'_{Train}/m$ and $K_{22}=$ $M_{Test}M'_{Test}/m$  then using the Woodbury matrix identity at the third step (\cite{petersen2008matrix}), we have \small \begin{align*}PEV(M_{Test})&= M_{Test}(M_{Train}M_{Train}+\lambda I)_{-1} M'_{Test}\\
&=M_{Test}(\lambda(\frac{M'_{Train}M_{Train}}{\lambda}+I))^{-1}M'_{test}\\
&=\frac{1}{\lambda}M_{Test}(I-M'_{Train}(M_{Train}M'_{Train}+\lambda I)^{-1}M_{Train})M'_{Test}\\
&= \frac{1}{\lambda}\left[M_{Test}M'_{Test}-M_{Test}M'_{Train}(M_{Train}M'_{Train}+\lambda I)^{-1}M_{Train}M'_{Test}\right]\\
&\propto K_{22} -K_{21}(K_{11}+m\lambda I)^{-1}K'_{21}.
\end{align*}
\normalsize

Therefore, maximizing average reliability is equivalent to minimizing the total $PEV^{Ridge}$ in (\ref{PEVRIDGE}), however since we would like to be evaluate many candidate training sets in the course of optimization we prefer the computationally efficient approximation in (\ref{PEVRIDGEAPPROX}). The scalar measure obtained by taking the trace of (\ref{PEVRIDGEAPPROX}) will be used to evaluate training populations subsequently.

The training selection optimization is a  combinatorial optimization problem. Genetic algorithms where a population of candidate solutions that are represented as binary strings of 0s and 1s is evolved toward better solutions. At each iteration of the algorithm a fitness function is used to evaluate and select the elite individuals and  subsequently the next population is formed from the elites by genetically motivated operations like crossover, mutation. Genetic algorithms are particularly suitable for optimization of combinatorial problems, therefore its our choice here. It should be noted that the solutions to the obtained by the genetic algorithm will usually be sub-optimal and different solutions can be obtained different starting points are used.

In the following section we will compare our training population selection scheme will be evaluated by fitting a semi-parametric mixed model (SPMM) (\cite{de2010semi,gianola2008reproducing}) using the genotypes and phenotypes in the training set and calculating the correlation of the test set phenotypes to the estimates based on this model.  In these mixed models genetic information in the form of a pedigree or markers are used to construct an additive relationship matrix that describes the similarity of line specific additive genetic effects. These models have been successfully used for predicting the breeding values in plants and animals.

A SPMM for the $n\times 1$ response vector $\by$ is expressed as 
\begin{equation}\label{eq:spmm} \by=X\beta+Z\bg+\be \end{equation} where $X$ is the $n\times p$ design matrix for the fixed effects, $\beta$ is a $p\times 1$ vector of fixed effects coefficients, $Z$ is the $n\times q$ design matrix for the random effects; the random effects $(\bg',\be')'$ are assumed to follow a multivariate normal distribution with mean $\bzero$ and covariance \[ \left( \begin{array}{cc}
\sigma^{2}_{g} K  & \bzero  \\
\bzero & \sigma^{2}_{e} I_{n} \end{array} \right)\] where $K$ is  a $q\times q$ relationship matrix. For fitting the mixed models we have developed and  utilized the EMMREML package (\cite{EMMREMLpackage})  which is available in R (\cite{team2005r}). The rest of the software was also programmed in R and are available in the supplementary files.

 An additive relationship matrix can be calculated from the centered scaled markers $M$ as $K=MM' /m.$ Given a similarity matrix $K$ the principal components used in our algorithm can be calculated from this matrix therefore the statistic in (\ref{PEVRIDGEAPPROX}) can also be used in these cases.   

\section*{Results}

Data sets of different origins are used for illustrations in this section. The Arabidopsis data set was published by Atwell et al. (2010) and is available at \url{https://cynin.gmi.oeaw.ac.at/home/resources/atpolydb/}.  The wheat data was downloaded from  \url{triticeaetoolbox.org}. The rice data was published in \cite{zhao2011genome} and was downloaded from  \url{http://www.ricediversity.org/data/}. These data sets are also available for download in the supplementary files.

 In order to evaluate the performance of the selection algorithm, we have  devised the following illustrations.

\begin{ex}
Arabidopsis data set consisted of genotypes of $199$ inbred lines along with observations on $107$ traits. Here we will report the result for 50 of these traits. 

For each trait first a test sample of size $n_{Test}=50$ was identified. From the remaining genotypes $n_{Train}=25,50,80$ were selected in the training population by random sampling or by the optimization method described in the previous section. The accuracies of the models were calculated by comparing the GEBVs with the observed phenotypes. This was repeated 30 times and the results are summarized in Figure \ref{fig:Arabidopsis}. At all sample sizes and for the vast majority of the traits the optimized samples improve accuracies as compared of the random samples. The difference is larger in general for smaller sample sizes and seems to decrease as the sample size increases.

\end{ex}

The accuracies of the genomic selection models tend to decrease as the training and test populations diverge. We claim that this can be partially remedied by optimizing training populations for the target population where the estimates are needed. The results from the next examples justify this claim.

\begin{ex}
5087 markers for 3975 elite wheat lines in the National Small Grains Collection (NSGC) were used for this example. In this experiment the thousand kernel weights were observed for non-overlapping subsets of the genotypes over five years (108 genotypes in year 2005, 416 in 2006, 281 in 2007,1358 in 2008 and 1896 in 2009).   We want to obtain the GEBVs of the genotypes for each of the years 2007 to 2009 from the genotypes that were observed before that year. The GEBVs for a random sample of $n_{Test}=200$ genotypes in the current year are estimated using by first a random sample and then an optimized sample of sizes $n_{Train}=100,300$ genotypes and phenotypes from the years preceding the test year. The experiment was repeated 30 times and the results are summarized with the box plots in Figure \ref{fig:WheatDataFiveYears}.The results are similar, models from optimized samples outperform the models from same size random samples and this difference decreases as the training sample size increases.

\end{ex}

In the next example,  we use a highly structured population and apply our population selection method in two different scenarios.
\begin{ex}
A diverse collection of 395 O. sativa (rice) accessions including both land races and elite varieties which represent the range of geographic and  genetic diversity of the species was used in this example. In addition to measurements for 36 continuous traits, genetic data on 40K SNPs were available for these 395 accessions. This data was first presented in \cite{zhao2011genome} and was also analyzed in \cite{wimmer2013genome}. We have selected five of these traits for our analysis, namely florets per panicle (FP),  panicle fertility, seed length (SL), seed weight (SW), seed surface area (SSA) and straighthead susceptability (SHS). For each of these traits a different subset of genotypes had the trait values.

In the first scenario, for each trait first a test sample of size $n_{Test}=100$ was identified. From the remaining genotypes $n_{Train}=25,50,100$ were selected in the training population by random sampling or by the optimization method described in the previous section. The accuracies of the models were calculated by comparing the GEBVs with the observed phenotypes. This was repeated 30 times and the results are summarized in Figure \ref{fig:Rice1}.

\end{ex}

Our last example is about evaluating the ability of estimating across clusters in a highly structured Maize data set. 

\begin{ex}
This data is given in \cite{romay2013comprehensive} and was also analyzed in \cite{wimmer2013genome}. 68120 markers on 2279 USA national inbred maize lines and their phenotypic means for degree days to silking compose the data set. 

We have first clustered the data into five clusters using the Euclidean distance matrix and the Ward's criterion for hierarchical clustering. The number of individuals in the resulting clusters were 1317 genotypes in the first cluster, 184 in second, 552 in third, 95 in forth and 131 in the fifth.  

From each of these clusters a test data set of size $n_{Test}=50$ was selected at random and a training population of size $n_{Train}=50, 100, 200$ genotypes from the remaining clusters were selected by random sampling or with the optimization scheme recommended in this article. The accuracies for estimating the observed trait values in each of these clusters were calculated for 30 independent replications and they are summarized in Figure \ref{fig:AmesData}.
Once again the optimized training sets outperform the random samples of the same size. 

\end{ex}

\section*{Conclusions}
In this article we have taken on the training selection problem and have shown by examples that incorporating information about the test set when available can improve the accuracies of prediction models. The approach we developed here is also computationally efficient. 
 
As seen from the examples in the previous section,the accuracy of the prediction models can be improved if the genotypes selected in the training population using our scheme especially when the required training sample size is small. By eliminating the irrelevant, outlier or influential individuals to enter into the model, and by ensuring that the a diverse training data set that adequately represent the test data set optimized training populations attain highly accurate models even when the training and test sets are not sampled from the same populations.

In the examples in previous section, we have selected the training populations separately for each trait. This was mainly because a different subset of genotypes were observed for different traits in the data sets. In practice however it would be satisfactory to select a single training population for all the traits with similar heritabilities because in the real setting phenotyping will follow this step and  procedure is robust to the choice of the shrinkage parameter $\lambda.$

We have discussed the training population problem in the context of the regression of continuous traits on the genotypes based on SPMMs. However, this approach can be used to obtain more accurate prediction models in different domains, i.e., in the general statistical learning domain. Our methods are useful for all high dimensional prediction problems where per individual cost of observing / analyzing the response variable is too high and a small number of training examples is sought and when the candidate data set is not representative of the test data set. 

Our results also indicate that genetic algorithm scheme adopted in this article is very efficient in finding a good solution in training population selection problem. However, there is no guarantee that the solutions found by this algorithm are the globally optimal solutions. Since the purpose of the article was to evaluate the overall improvement over many replications of the same experiments it was not feasible for us to start the genetic algorithm at different starting points but when it is affordable it would be safer to do so. 

A dynamic model building approach might be more suitable when the genotypes in the test set are highly structured. It might be possible to improve accuracies using a different model for different parts of the tests set built on only the genotypes selected by the training population selection algorithm. Another approach we have not tried, but worth additional inquiry, is to estimate each test point with a different model.

\begin{backmatter}

\section*{Competing interests}
  The authors declare that they have no competing interests.

\section*{Author's contributions}
  Deniz Akdemir (Corresponding Author): Idea, text \& programs.\\

\section*{Acknowledgments}
This research was supported by the USDA-NIFA-AFRI Triticeae Coordinated Agricultural Project, award number 2011-68002-30029.


\bibliographystyle{bmc-mathphys} 
\bibliography{PEVmean.bib}  


\begin{thebibliography}{13}
\ifx \bisbn   \undefined \def \bisbn  #1{ISBN #1}\fi
\ifx \binits  \undefined \def \binits#1{#1}\fi
\ifx \bauthor  \undefined \def \bauthor#1{#1}\fi
\ifx \batitle  \undefined \def \batitle#1{#1}\fi
\ifx \bjtitle  \undefined \def \bjtitle#1{#1}\fi
\ifx \bvolume  \undefined \def \bvolume#1{\textbf{#1}}\fi
\ifx \byear  \undefined \def \byear#1{#1}\fi
\ifx \bissue  \undefined \def \bissue#1{#1}\fi
\ifx \bfpage  \undefined \def \bfpage#1{#1}\fi
\ifx \blpage  \undefined \def \blpage #1{#1}\fi
\ifx \burl  \undefined \def \burl#1{\textsf{#1}}\fi
\ifx \doiurl  \undefined \def \doiurl#1{\textsf{#1}}\fi
\ifx \betal  \undefined \def \betal{\textit{et al.}}\fi
\ifx \binstitute  \undefined \def \binstitute#1{#1}\fi
\ifx \binstitutionaled  \undefined \def \binstitutionaled#1{#1}\fi
\ifx \bctitle  \undefined \def \bctitle#1{#1}\fi
\ifx \beditor  \undefined \def \beditor#1{#1}\fi
\ifx \bpublisher  \undefined \def \bpublisher#1{#1}\fi
\ifx \bbtitle  \undefined \def \bbtitle#1{#1}\fi
\ifx \bedition  \undefined \def \bedition#1{#1}\fi
\ifx \bseriesno  \undefined \def \bseriesno#1{#1}\fi
\ifx \blocation  \undefined \def \blocation#1{#1}\fi
\ifx \bsertitle  \undefined \def \bsertitle#1{#1}\fi
\ifx \bsnm \undefined \def \bsnm#1{#1}\fi
\ifx \bsuffix \undefined \def \bsuffix#1{#1}\fi
\ifx \bparticle \undefined \def \bparticle#1{#1}\fi
\ifx \barticle \undefined \def \barticle#1{#1}\fi
\ifx \bconfdate \undefined \def \bconfdate #1{#1}\fi
\ifx \botherref \undefined \def \botherref #1{#1}\fi
\ifx \url \undefined \def \url#1{\textsf{#1}}\fi
\ifx \bchapter \undefined \def \bchapter#1{#1}\fi
\ifx \bbook \undefined \def \bbook#1{#1}\fi
\ifx \bcomment \undefined \def \bcomment#1{#1}\fi
\ifx \oauthor \undefined \def \oauthor#1{#1}\fi
\ifx \citeauthoryear \undefined \def \citeauthoryear#1{#1}\fi
\ifx \endbibitem  \undefined \def \endbibitem {}\fi
\ifx \bconflocation  \undefined \def \bconflocation#1{#1}\fi
\ifx \arxivurl  \undefined \def \arxivurl#1{\textsf{#1}}\fi
\csname PreBibitemsHook\endcsname

\bibitem{vanraden2008}
\begin{barticle}
\bauthor{\bsnm{VanRaden}, \binits{P.}}:
\batitle{Efficient methods to compute genomic predictions}.
\bjtitle{Journal of dairy science}
\bvolume{91}(\bissue{11}),
\bfpage{4414}--\blpage{4423}
(\byear{2008})
\end{barticle}
\endbibitem

\bibitem{henderson1975best}
\begin{botherref}
\oauthor{\bsnm{Henderson}, \binits{C.R.}}:
Best linear unbiased estimation and prediction under a selection model.
Biometrics,
423--447
(1975)
\end{botherref}
\endbibitem

\bibitem{laloe1996considerations}
\begin{barticle}
\bauthor{\bsnm{Lalo{\"e}}, \binits{D.}},
\bauthor{\bsnm{Phocas}, \binits{F.}},
\bauthor{\bsnm{M{\'e}nissier}, \binits{F.}}:
\batitle{Considerations on measures of precision and connectedness in mixed
  linear models of genetic evaluation}.
\bjtitle{Genetics Selection Evolution}
\bvolume{28}(\bissue{4}),
\bfpage{359}--\blpage{378}
(\byear{1996})
\end{barticle}
\endbibitem

\bibitem{rincent2012maximizing}
\begin{barticle}
\bauthor{\bsnm{Rincent}, \binits{R.}},
\bauthor{\bsnm{Lalo{\"e}}, \binits{D.}},
\bauthor{\bsnm{Nicolas}, \binits{S.}},
\bauthor{\bsnm{Altmann}, \binits{T.}},
\bauthor{\bsnm{Brunel}, \binits{D.}},
\bauthor{\bsnm{Revilla}, \binits{P.}},
\bauthor{\bsnm{Rodriguez}, \binits{V.M.}},
\bauthor{\bsnm{Moreno-Gonzalez}, \binits{J.}},
\bauthor{\bsnm{Melchinger}, \binits{A.}},
\bauthor{\bsnm{Bauer}, \binits{E.}}, \betal:
\batitle{Maximizing the reliability of genomic selection by optimizing the
  calibration set of reference individuals: Comparison of methods in two
  diverse groups of maize inbreds (zea mays l.)}.
\bjtitle{Genetics}
\bvolume{192}(\bissue{2}),
\bfpage{715}--\blpage{728}
(\byear{2012})
\end{barticle}
\endbibitem

\bibitem{box1959basis}
\begin{barticle}
\bauthor{\bsnm{Box}, \binits{G.E.}},
\bauthor{\bsnm{Draper}, \binits{N.R.}}:
\batitle{A basis for the selection of a response surface design}.
\bjtitle{Journal of the American Statistical Association}
\bvolume{54}(\bissue{287}),
\bfpage{622}--\blpage{654}
(\byear{1959})
\end{barticle}
\endbibitem

\bibitem{petersen2008matrix}
\begin{botherref}
\oauthor{\bsnm{Petersen}, \binits{K.B.}},
\oauthor{\bsnm{Pedersen}, \binits{M.S.}}:
The matrix cookbook.
Technical University of Denmark,
7--15
(2008)
\end{botherref}
\endbibitem

\bibitem{de2010semi}
\begin{barticle}
\bauthor{\bparticle{de} \bsnm{Los~Campos}, \binits{G.}},
\bauthor{\bsnm{Gianola}, \binits{D.}},
\bauthor{\bsnm{Rosa}, \binits{G.J.}},
\bauthor{\bsnm{Weigel}, \binits{K.A.}},
\bauthor{\bsnm{Crossa}, \binits{J.}}:
\batitle{Semi-parametric genomic-enabled prediction of genetic values using
  reproducing kernel hilbert spaces methods}.
\bjtitle{Genetics Research}
\bvolume{92}(\bissue{04}),
\bfpage{295}--\blpage{308}
(\byear{2010})
\end{barticle}
\endbibitem

\bibitem{gianola2008reproducing}
\begin{barticle}
\bauthor{\bsnm{Gianola}, \binits{D.}},
\bauthor{\bparticle{van} \bsnm{Kaam}, \binits{J.B.}}:
\batitle{Reproducing kernel hilbert spaces regression methods for genomic
  assisted prediction of quantitative traits}.
\bjtitle{Genetics}
\bvolume{178}(\bissue{4}),
\bfpage{2289}--\blpage{2303}
(\byear{2008})
\end{barticle}
\endbibitem

\bibitem{EMMREMLpackage}
\begin{botherref}
\oauthor{\bsnm{Akdemir}, \binits{D.}}:
R Package ''EMMREML''
(2014)
\end{botherref}
\endbibitem

\bibitem{team2005r}
\begin{botherref}
\oauthor{\bsnm{RCoreTeam}}:
R: A language and environment for statistical computing.
sn
(2005)
\end{botherref}
\endbibitem

\bibitem{zhao2011genome}
\begin{barticle}
\bauthor{\bsnm{Zhao}, \binits{K.}},
\bauthor{\bsnm{Tung}, \binits{C.-W.}},
\bauthor{\bsnm{Eizenga}, \binits{G.C.}},
\bauthor{\bsnm{Wright}, \binits{M.H.}},
\bauthor{\bsnm{Ali}, \binits{M.L.}},
\bauthor{\bsnm{Price}, \binits{A.H.}},
\bauthor{\bsnm{Norton}, \binits{G.J.}},
\bauthor{\bsnm{Islam}, \binits{M.R.}},
\bauthor{\bsnm{Reynolds}, \binits{A.}},
\bauthor{\bsnm{Mezey}, \binits{J.}}, \betal:
\batitle{Genome-wide association mapping reveals a rich genetic architecture of
  complex traits in oryza sativa}.
\bjtitle{Nature communications}
\bvolume{2},
\bfpage{467}
(\byear{2011})
\end{barticle}
\endbibitem

\bibitem{wimmer2013genome}
\begin{barticle}
\bauthor{\bsnm{Wimmer}, \binits{V.}},
\bauthor{\bsnm{Lehermeier}, \binits{C.}},
\bauthor{\bsnm{Albrecht}, \binits{T.}},
\bauthor{\bsnm{Auinger}, \binits{H.-J.}},
\bauthor{\bsnm{Wang}, \binits{Y.}},
\bauthor{\bsnm{Sch{\"o}n}, \binits{C.-C.}}:
\batitle{Genome-wide prediction of traits with different genetic architecture
  through efficient variable selection}.
\bjtitle{Genetics}
\bvolume{195}(\bissue{2}),
\bfpage{573}--\blpage{587}
(\byear{2013})
\end{barticle}
\endbibitem

\bibitem{romay2013comprehensive}
\begin{barticle}
\bauthor{\bsnm{Romay}, \binits{M.C.}},
\bauthor{\bsnm{Millard}, \binits{M.J.}},
\bauthor{\bsnm{Glaubitz}, \binits{J.C.}},
\bauthor{\bsnm{Peiffer}, \binits{J.A.}},
\bauthor{\bsnm{Swarts}, \binits{K.L.}},
\bauthor{\bsnm{Casstevens}, \binits{T.M.}},
\bauthor{\bsnm{Elshire}, \binits{R.J.}},
\bauthor{\bsnm{Acharya}, \binits{C.B.}},
\bauthor{\bsnm{Mitchell}, \binits{S.E.}},
\bauthor{\bsnm{Flint-Garcia}, \binits{S.A.}}, \betal:
\batitle{Comprehensive genotyping of the usa national maize inbred seed bank}.
\bjtitle{Genome biology}
\bvolume{14}(\bissue{6}),
\bfpage{55}
(\byear{2013})
\end{barticle}
\endbibitem

\end{thebibliography}

\newcommand{\BMCxmlcomment}[1]{}

\BMCxmlcomment{

<refgrp>

<bibl id="B1">
  <title><p>Efficient methods to compute genomic predictions</p></title>
  <aug>
    <au><snm>VanRaden</snm><fnm>PM</fnm></au>
  </aug>
  <source>Journal of dairy science</source>
  <publisher>Elsevier</publisher>
  <pubdate>2008</pubdate>
  <volume>91</volume>
  <issue>11</issue>
  <fpage>4414</fpage>
  <lpage>-4423</lpage>
</bibl>

<bibl id="B2">
  <title><p>Best linear unbiased estimation and prediction under a selection
  model</p></title>
  <aug>
    <au><snm>Henderson</snm><fnm>CR</fnm></au>
  </aug>
  <source>Biometrics</source>
  <publisher>JSTOR</publisher>
  <pubdate>1975</pubdate>
  <fpage>423</fpage>
  <lpage>-447</lpage>
</bibl>

<bibl id="B3">
  <title><p>Considerations on measures of precision and connectedness in mixed
  linear models of genetic evaluation</p></title>
  <aug>
    <au><snm>Lalo{\"e}</snm><fnm>D</fnm></au>
    <au><snm>Phocas</snm><fnm>F</fnm></au>
    <au><snm>M{\'e}nissier</snm><fnm>F</fnm></au>
  </aug>
  <source>Genetics Selection Evolution</source>
  <publisher>EDP Sciences</publisher>
  <pubdate>1996</pubdate>
  <volume>28</volume>
  <issue>4</issue>
  <fpage>359</fpage>
  <lpage>-378</lpage>
</bibl>

<bibl id="B4">
  <title><p>Maximizing the Reliability of Genomic Selection by Optimizing the
  Calibration Set of Reference Individuals: Comparison of Methods in Two
  Diverse Groups of Maize Inbreds (Zea mays L.)</p></title>
  <aug>
    <au><snm>Rincent</snm><fnm>R</fnm></au>
    <au><snm>Lalo{\"e}</snm><fnm>D</fnm></au>
    <au><snm>Nicolas</snm><fnm>S</fnm></au>
    <au><snm>Altmann</snm><fnm>T</fnm></au>
    <au><snm>Brunel</snm><fnm>D</fnm></au>
    <au><snm>Revilla</snm><fnm>P</fnm></au>
    <au><snm>Rodriguez</snm><fnm>VM</fnm></au>
    <au><snm>Moreno Gonzalez</snm><fnm>J</fnm></au>
    <au><snm>Melchinger</snm><fnm>A</fnm></au>
    <au><snm>Bauer</snm><fnm>E</fnm></au>
    <au><cnm>others</cnm></au>
  </aug>
  <source>Genetics</source>
  <publisher>Genetics Society of America</publisher>
  <pubdate>2012</pubdate>
  <volume>192</volume>
  <issue>2</issue>
  <fpage>715</fpage>
  <lpage>-728</lpage>
</bibl>

<bibl id="B5">
  <title><p>A basis for the selection of a response surface design</p></title>
  <aug>
    <au><snm>Box</snm><fnm>GE</fnm></au>
    <au><snm>Draper</snm><fnm>NR</fnm></au>
  </aug>
  <source>Journal of the American Statistical Association</source>
  <publisher>Taylor \& Francis</publisher>
  <pubdate>1959</pubdate>
  <volume>54</volume>
  <issue>287</issue>
  <fpage>622</fpage>
  <lpage>-654</lpage>
</bibl>

<bibl id="B6">
  <title><p>The matrix cookbook</p></title>
  <aug>
    <au><snm>Petersen</snm><fnm>KB</fnm></au>
    <au><snm>Pedersen</snm><fnm>MS</fnm></au>
  </aug>
  <source>Technical University of Denmark</source>
  <pubdate>2008</pubdate>
  <fpage>7</fpage>
  <lpage>-15</lpage>
</bibl>

<bibl id="B7">
  <title><p>Semi-parametric genomic-enabled prediction of genetic values using
  reproducing kernel Hilbert spaces methods</p></title>
  <aug>
    <au><snm>Los Campos</snm><fnm>G</fnm></au>
    <au><snm>Gianola</snm><fnm>D</fnm></au>
    <au><snm>Rosa</snm><fnm>GJ</fnm></au>
    <au><snm>Weigel</snm><fnm>KA</fnm></au>
    <au><snm>Crossa</snm><fnm>J</fnm></au>
  </aug>
  <source>Genetics Research</source>
  <publisher>Cambridge Univ Press</publisher>
  <pubdate>2010</pubdate>
  <volume>92</volume>
  <issue>04</issue>
  <fpage>295</fpage>
  <lpage>-308</lpage>
</bibl>

<bibl id="B8">
  <title><p>Reproducing kernel Hilbert spaces regression methods for genomic
  assisted prediction of quantitative traits</p></title>
  <aug>
    <au><snm>Gianola</snm><fnm>D</fnm></au>
    <au><snm>Kaam</snm><fnm>JB</fnm></au>
  </aug>
  <source>Genetics</source>
  <publisher>Genetics Soc America</publisher>
  <pubdate>2008</pubdate>
  <volume>178</volume>
  <issue>4</issue>
  <fpage>2289</fpage>
  <lpage>-2303</lpage>
</bibl>

<bibl id="B9">
  <title><p>R Package ''EMMREML''</p></title>
  <aug>
    <au><snm>Akdemir</snm><fnm>D</fnm></au>
  </aug>
  <pubdate>2014</pubdate>
</bibl>

<bibl id="B10">
  <title><p>R: A language and environment for statistical computing</p></title>
  <aug>
    <au><cnm>RCoreTeam</cnm></au>
  </aug>
  <source>R foundation for Statistical Computing</source>
  <publisher>sn</publisher>
  <pubdate>2005</pubdate>
</bibl>

<bibl id="B11">
  <title><p>Genome-wide association mapping reveals a rich genetic architecture
  of complex traits in Oryza sativa</p></title>
  <aug>
    <au><snm>Zhao</snm><fnm>K</fnm></au>
    <au><snm>Tung</snm><fnm>CW</fnm></au>
    <au><snm>Eizenga</snm><fnm>GC</fnm></au>
    <au><snm>Wright</snm><fnm>MH</fnm></au>
    <au><snm>Ali</snm><fnm>ML</fnm></au>
    <au><snm>Price</snm><fnm>AH</fnm></au>
    <au><snm>Norton</snm><fnm>GJ</fnm></au>
    <au><snm>Islam</snm><fnm>MR</fnm></au>
    <au><snm>Reynolds</snm><fnm>A</fnm></au>
    <au><snm>Mezey</snm><fnm>J</fnm></au>
    <au><cnm>others</cnm></au>
  </aug>
  <source>Nature communications</source>
  <publisher>Nature Publishing Group</publisher>
  <pubdate>2011</pubdate>
  <volume>2</volume>
  <fpage>467</fpage>
</bibl>

<bibl id="B12">
  <title><p>Genome-wide prediction of traits with different genetic
  architecture through efficient variable selection</p></title>
  <aug>
    <au><snm>Wimmer</snm><fnm>V</fnm></au>
    <au><snm>Lehermeier</snm><fnm>C</fnm></au>
    <au><snm>Albrecht</snm><fnm>T</fnm></au>
    <au><snm>Auinger</snm><fnm>HJ</fnm></au>
    <au><snm>Wang</snm><fnm>Y</fnm></au>
    <au><snm>Sch{\"o}n</snm><fnm>CC</fnm></au>
  </aug>
  <source>Genetics</source>
  <publisher>Genetics Soc America</publisher>
  <pubdate>2013</pubdate>
  <volume>195</volume>
  <issue>2</issue>
  <fpage>573</fpage>
  <lpage>-587</lpage>
</bibl>

<bibl id="B13">
  <title><p>Comprehensive genotyping of the USA national maize inbred seed
  bank</p></title>
  <aug>
    <au><snm>Romay</snm><fnm>MC</fnm></au>
    <au><snm>Millard</snm><fnm>MJ</fnm></au>
    <au><snm>Glaubitz</snm><fnm>JC</fnm></au>
    <au><snm>Peiffer</snm><fnm>JA</fnm></au>
    <au><snm>Swarts</snm><fnm>KL</fnm></au>
    <au><snm>Casstevens</snm><fnm>TM</fnm></au>
    <au><snm>Elshire</snm><fnm>RJ</fnm></au>
    <au><snm>Acharya</snm><fnm>CB</fnm></au>
    <au><snm>Mitchell</snm><fnm>SE</fnm></au>
    <au><snm>Flint Garcia</snm><fnm>SA</fnm></au>
    <au><cnm>others</cnm></au>
  </aug>
  <source>Genome biology</source>
  <publisher>BioMed Central Ltd</publisher>
  <pubdate>2013</pubdate>
  <volume>14</volume>
  <issue>6</issue>
  <fpage>R55</fpage>
</bibl>

</refgrp>
} 

\section*{Figures}
  
\begin{figure}
\centering
      \includegraphics[width=.8]{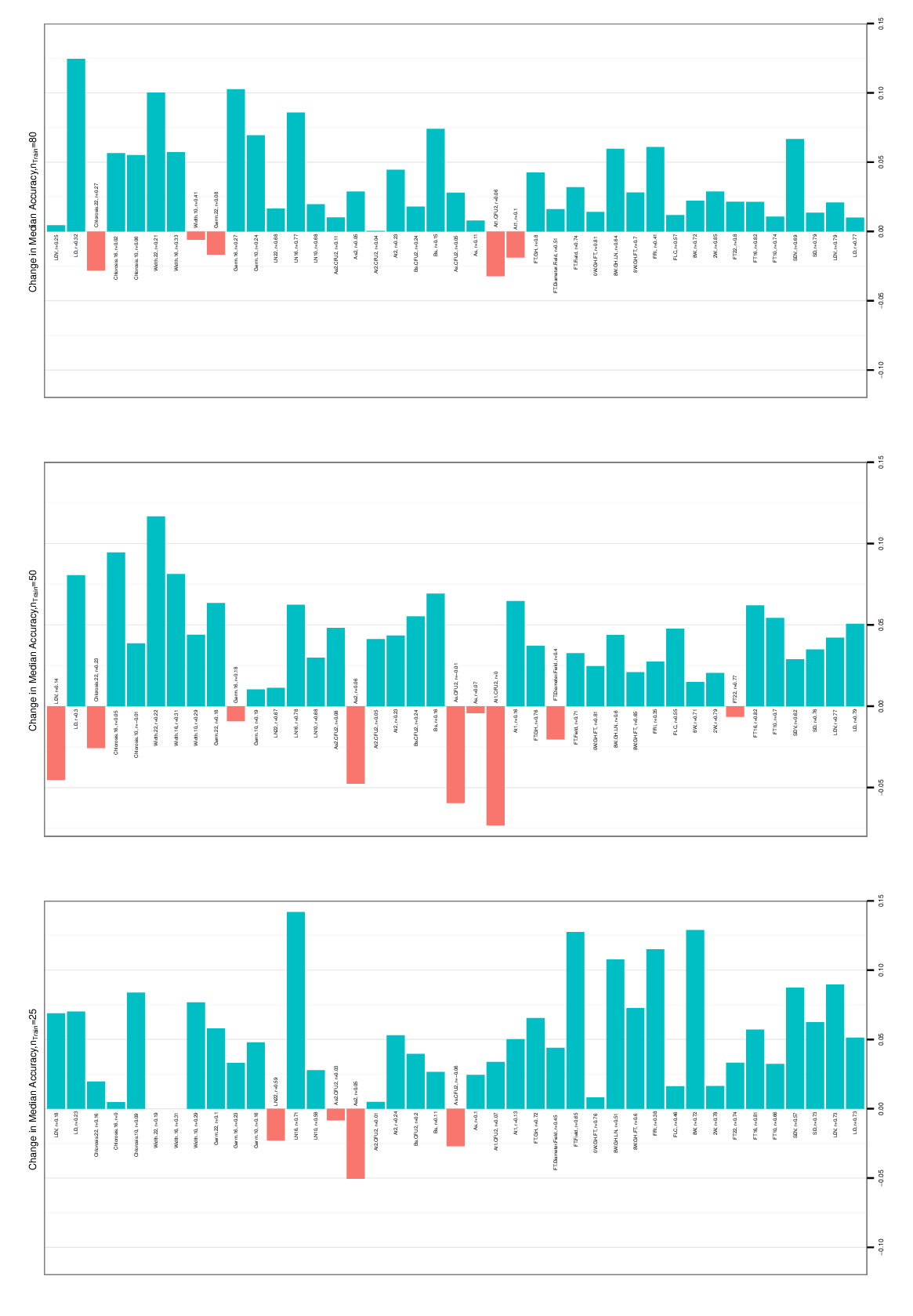}
\caption[Arabidopsis Data]{The difference between the accuracies of the models trained on optimized populations versus random samples. Positive values indicate the cases where the optimized population performed better as compared to the random sample. The median accuracies of the optimized sample for the traits  are also available by the corresponding bar.}
\label{fig:Arabidopsis}
\end{figure}

\begin{figure}
\centering
      \includegraphics[width=.8]{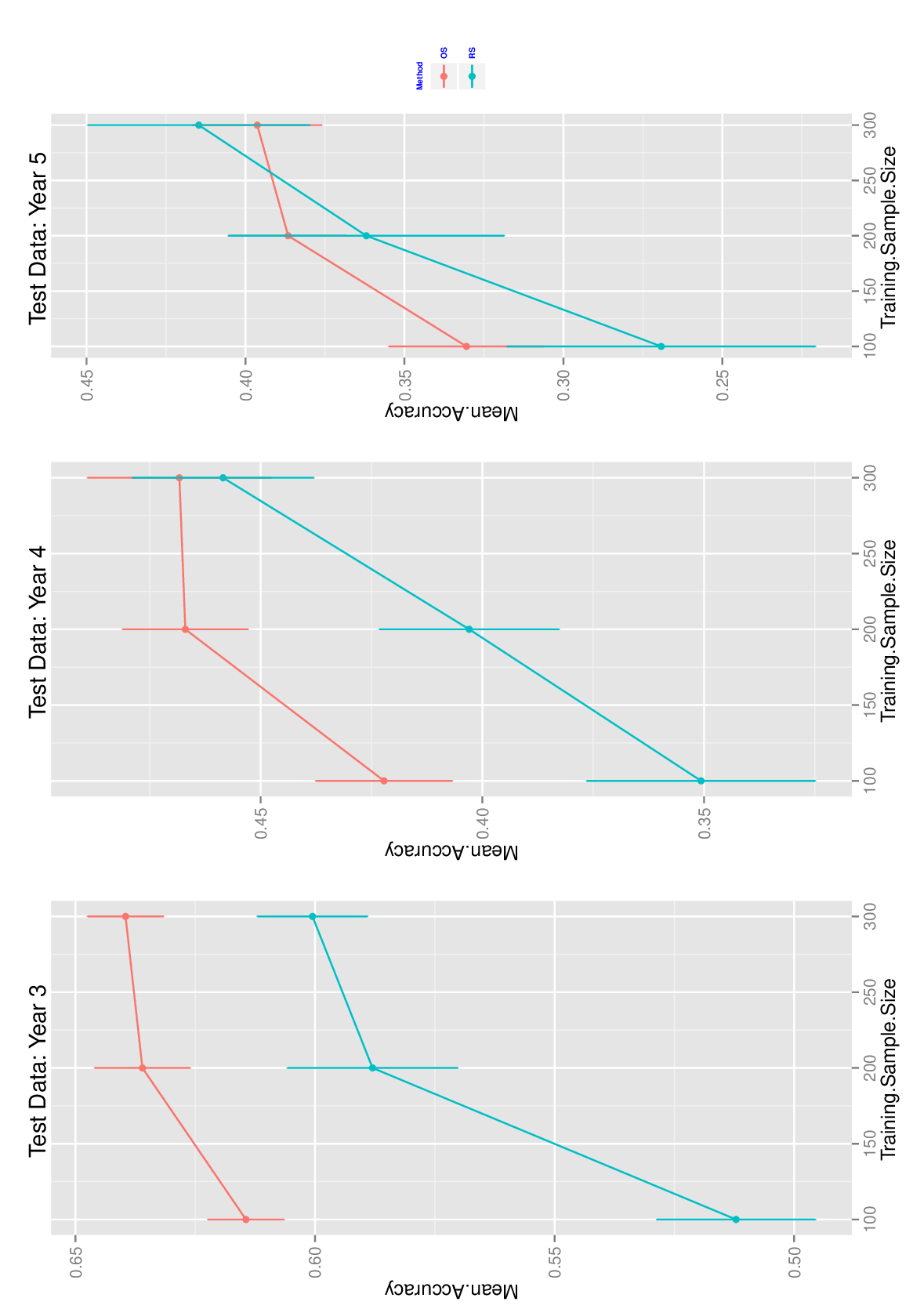}
\caption[Wheat Data]{The comparisons of the mean accuracies (measured by correlation) when the test data set is selected from years 2007 through 2009 for different training sample sizes. For each of these cases the training set was selected from the genotypes in the years preceding the test year.}
\label{fig:WheatDataFiveYears}
\end{figure}

\begin{figure}
\centering
      \includegraphics[width=.8]{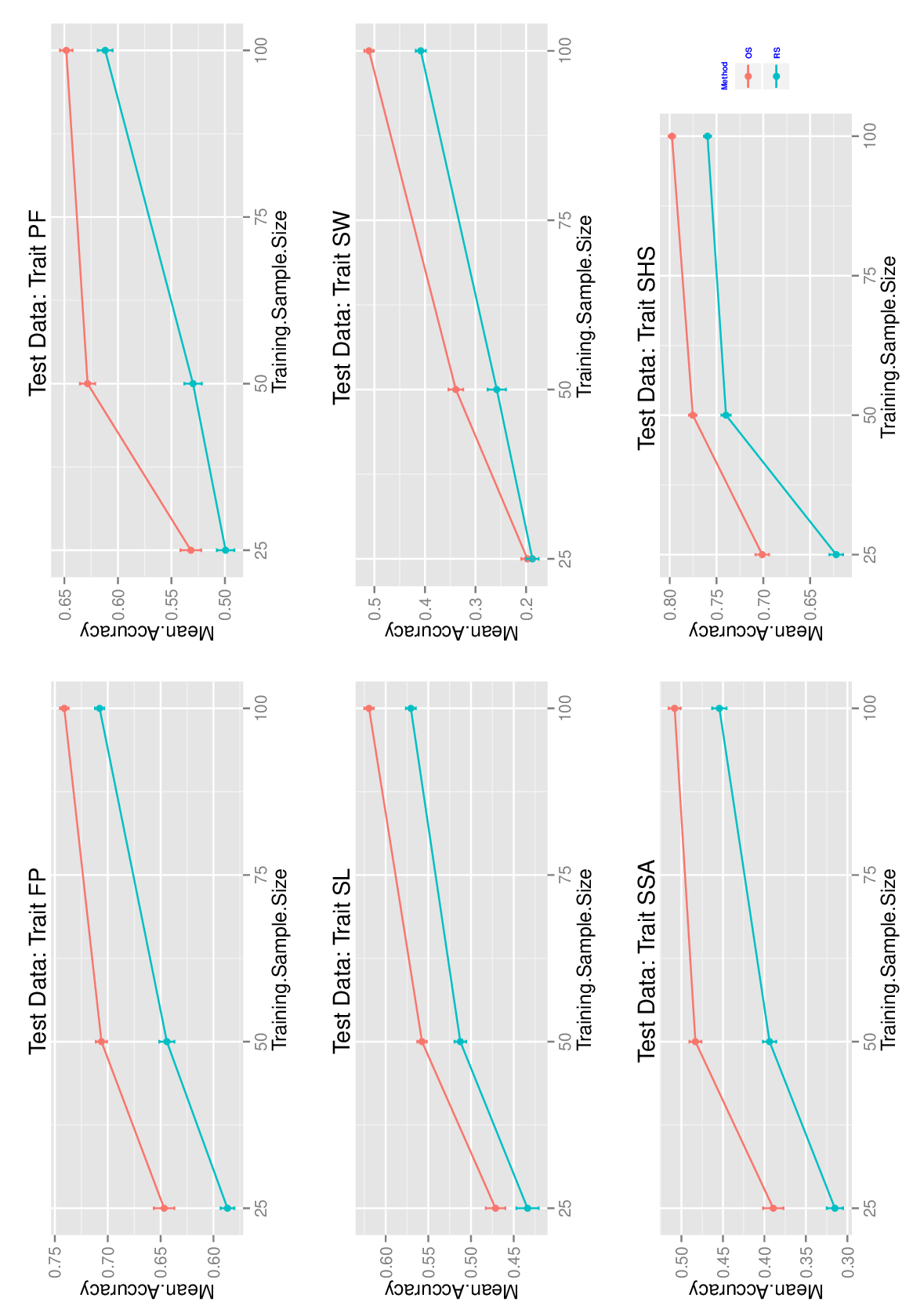}
\caption[Rice Data]{The comparisons of mean accuracies (measured by correlation) for the traits florets per panicle (FP),  panicle fertility, seed length (SL), seed weight (SW), seed surface area (SSA) and straighthead susceptability (SHS) for different training sample sizes. Optimized samples outperform random samples almost exclusively.}
\label{fig:Rice1}
\end{figure}

\begin{figure}
\centering
      \includegraphics[width=.8]{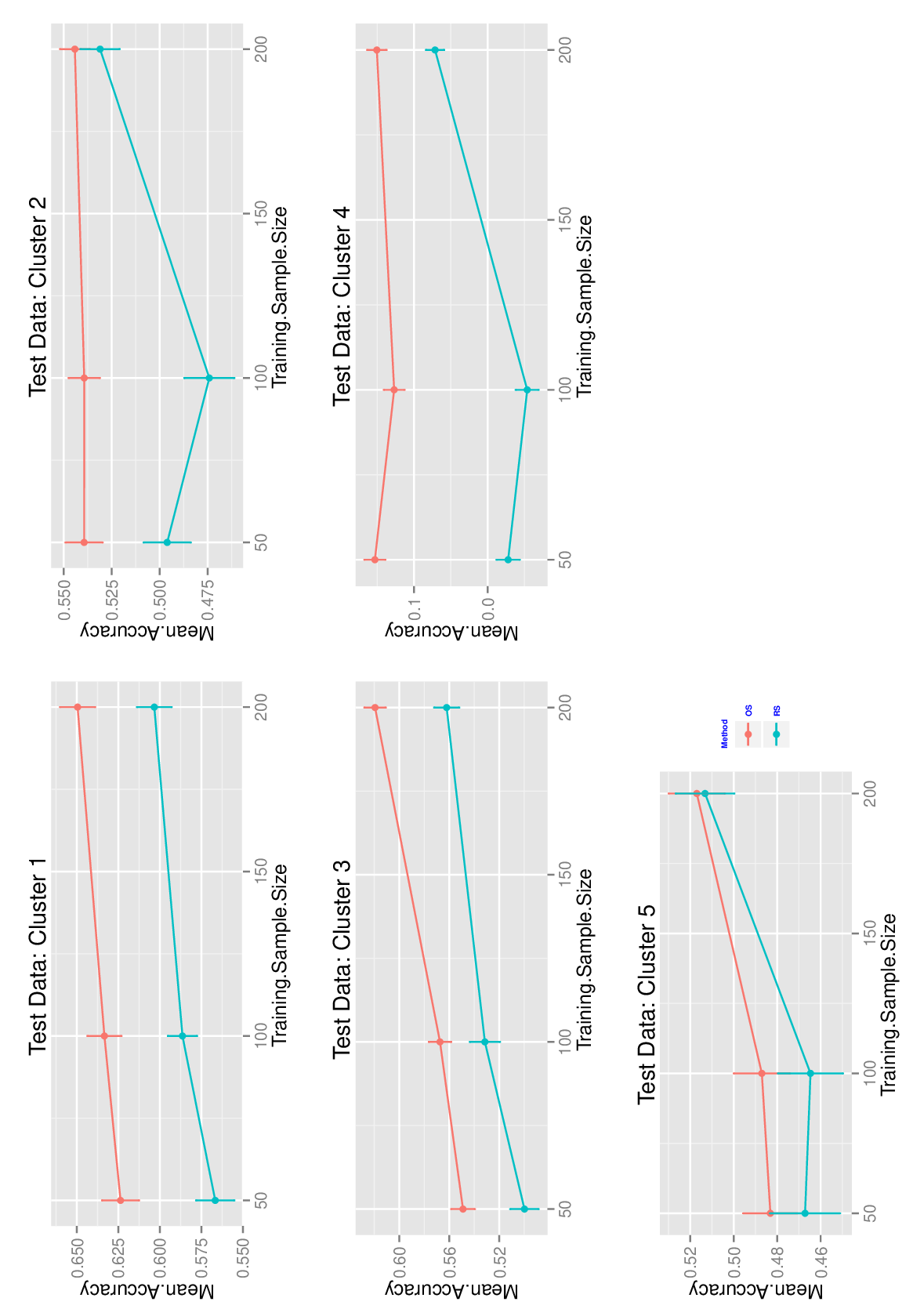}
\caption[Maize Data]{The comparisons of the accuracies  for prediction across clusters in the highly structured Maize data set. For a test data set of size $n_{Test}=50$ was selected at random in a particular cluster and a training population of size $n_{Train}=50, 100, 200$ genotypes was selected from the remaining clusters. The accuracies vary significantly from cluster to cluster however the optimized training set performs better on average.}
\label{fig:AmesData}
\end{figure}




\section*{Additional Files}
  \subsection*{Additional file --- R programs and the data sets. }
    trainingpopulationselection.zip 

\end{backmatter}
\end{document}